\documentclass{jpsj-suppl}

\usepackage{cite}

\makeatletter
 \DeclareRobustCommand\cite{\unskip
\@ifnextchar[{\@tempswatrue\@citex}{\@tempswafalse\@citex[]}}
 \def\@cite#1#2{$^{\hbox{\scriptsize{#1\if@tempswa , #2\fi})}}$}
 \def\@biblabel#1{#1)}
\makeatother

\usepackage{txfonts} 
\usepackage{wrapfig}

\title{Single crystal growth and physical properties of SrFe$_{2}$(As$_{1-x}$P$_{x}$)$_{2}$}

\author{Tatsuya \textsc{Kobayashi}$^{1}$, Shigeki \textsc{Miyasaka}$^{1, 2}$, Setsuko \textsc{Tajima}$^{1, 2}$}

\inst{$^{1}$Department of Physics, Graduate School of Science, Osaka University, Osaka 560-0043, Japan \\
$^{2}$JST, Transformative Research-Project on Iron Pnictides (TRIP), Tokyo 102-0075, Japan}

\email{kobayashi@tsurugi.phys.sci.osaka-u.ac.jp}

\abst{We report a crystal growth and physical properties of SrFe$_{2}$(As$_{1-x}$P$_{x}$)$_{2}$. The single crystals for various $x$s were grown by a self flux method. For $x = 0.35$, $T_c$ reaches the maximum value of 30\,K and the electrical resistivity $\rho$($T$) shows $T$-linear dependence. 
As $x$ increases, $T_{c}$ decreases and $\rho$($T$) changes to $T^2$-behavior, indicating a standard Fermi liquid. 
These results suggest that a magnetic quantum critical point exists around $x=0.35$. }

\kword{Ternary ThCr$_{2}$Si$_{2}$ type iron-pnictide, superconductors, synthesis of single crystal.}

\begin{document}
\maketitle

\section{INTRODUCTION}

Iron pnictide superconductor LaFeAsO$_{1-x}$F$_{x}$ was discovered in 2008, which shows $T_c$ = 26\,K\cite{1}. 
Soon after the discovery, $R$FeAsO$_{1-x}$F$_{x}$ ($R=$ Ce, Pr, Sm, Nd) was found and the maximum $T_c$ reached 55\,K. 
Moreover, new iron pnictide or chalcogenide superconductors with different crystal structure such as Ba$_{1-x}$K$_{x}$Fe$_{2}$As$_{2}$, 
LiFeAs, FeSe$_{1-x}$ and KFe$_{2}$Se$_{2}$ were reported one after another, but the superconducting mechanism is still unclear despite many intensive researches.

Superconductivity in $A$Fe$_{2}$As$_{2}$ ($A$ = Ba, Sr, Ca, Eu, so called $A1$22 system) is induced by hole-doping (K substitution for $A$) and electron-doping (Co for Fe). 
Additionally, isovalent substitution (P for As) also induces superconductivity. 
This system, particularly BaFe$_{2}$(As$_{1-x}$P$_{x}$)$_{2}$, attracts attention in terms of magnetic quantum criticality and the nodal superconducting gap feature \cite{2} in contrast to a full gap for most of the other iron based superconductors\cite{5,6}. 
Compared with many investigation for BaFe$_{2}$(As$_{1-x}$P$_{x}$)$_{2}$, the study of the related system, SrFe$_{2}$(As$_{1-x}$P$_{x}$)$_{2}$ is never reported except the polycrystal study\cite{4}, thus the study with single crystals is needed to elucidate the superconducting mechanism of P substituted $A$122 system.

In this study, we synthesized  single crystals of SrFe$_{2}$(As$_{1-x}$P$_{x}$)$_{2}$ and measured the physical properties to clarify a phase diagram and anomalous resistivity behaviors in the vicinity of magnetic quantum critical point.

\section{EXPERIMENTAL}

$A$Fe$_{2}$As$_{2}$ can be synthesized by several flux methods.
In this study, SrFe$_{2}$As$_{2}$ was synthesized with Sn flux method. 
Sr chunks, FeAs and Sn were loaded in an alumina crucible according to the ratio of Sr+2FeAs: Sn = 1: 25-40. 
The alumina crucible in a sealed silica tube was heated up to 1020\,\degC, kept for 12\,hours, and then cooled down to 600\,\degC~over 122\,hours. Sn flux was removed by using centrifuge. Plate like single crystals with typical size of 4~4~0.5\,mm$^{3}$ were obtained.

\begin{wrapfigure}{r}{45mm}
\begin{center}
\includegraphics[width=4cm]{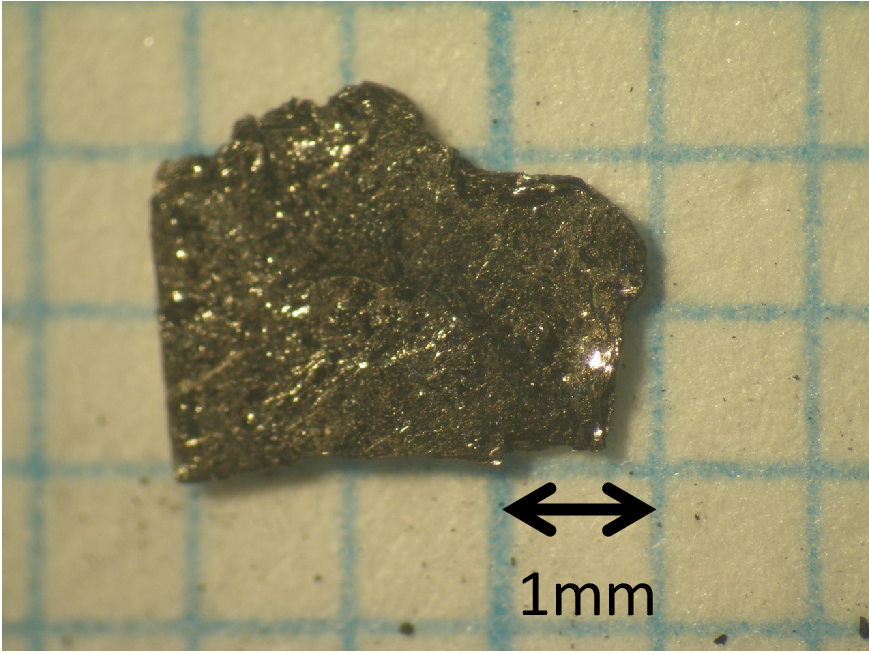}
\caption{(Color online)
Photograph of single crystal of SrFe$_{2}$(As$_{0.65}$P$_{0.35}$)$_{2}$.}
\label{crystal}
\end{center}
\end{wrapfigure}

On the other hand, SrFe$_{2}$(As$_{1-x}$P$_{x}$)$_{2}$ could not be obtained by a Sn flux or a self flux method using excess FeAs. 
So we grew single crystals of SrFe$_{2}$(As$_{1-x}$P$_{x}$)$_2$  from stoichiometric mixtures of Sr, FeAs, and FeP powders placed in an alumina crucible, sealed in a silica tube with Ar gas of 0.2\,bar at room temperature to prevent Sr from evaporating. It was heated up to 1230-1300\,\degC~relatively higher temperature than the case of crystal growth of Co substituted systems, kept for 12\,hours, and then slowly cooled down to 1050\,\degC~at the rate of 1-2\,\degC/h. 
Plate-like crystals, typical size of 1~1~0.13\,mm$^{3}$ were extracted (Fig. \ref{crystal}).The crystal size tends to become smaller as $x$ increases.

The electrical resistivity was measured by a standard four-probe method and the magnetic susceptibility was measured by a magnetic property measurement system (MPMS) of Quantum Design Company.

\section{RESULT AND DISCUSSION}

\begin{figure}
\begin{center}
\includegraphics[width=8cm]{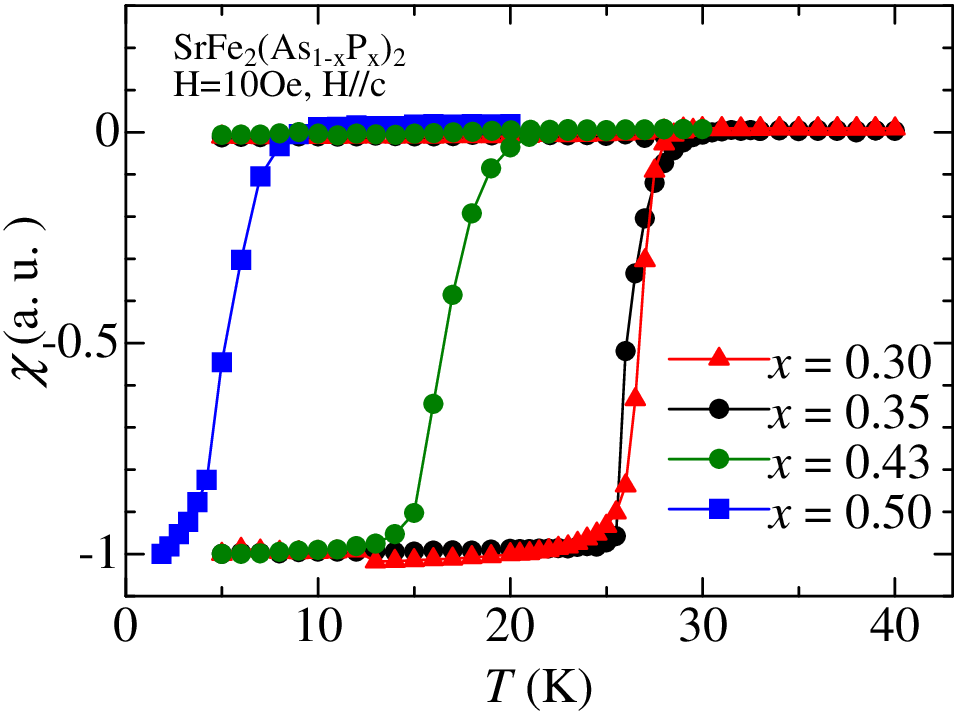}
\caption{(Color online)
Temperature dependence of magnetic susceptibility $\chi$ of SrFe$_{2}$(As$_{1-x}$P$_{x}$)$_{2}$ in magnetic field of $x=0.30,\,0.35,\,0.43,\,0.50$.
Measurements were performed the field-cooled and zero-field-cooled process at 10\,Oe with $H \,|| \,c$.}
\label{magnet}
\end{center}
\end{figure}

Figure \ref{magnet} presents the temperature dependent magnetic susceptibility in 10\,Oe, normalized to their lowest zero-field-cooled values. 
In zero-field-cooled data, there is a clear drop at the temperature associated with superconductivity. The field-cooled susceptibility data manifest clear Meisner effect. 

\begin{figure}
\begin{center}
\includegraphics[width=8cm]{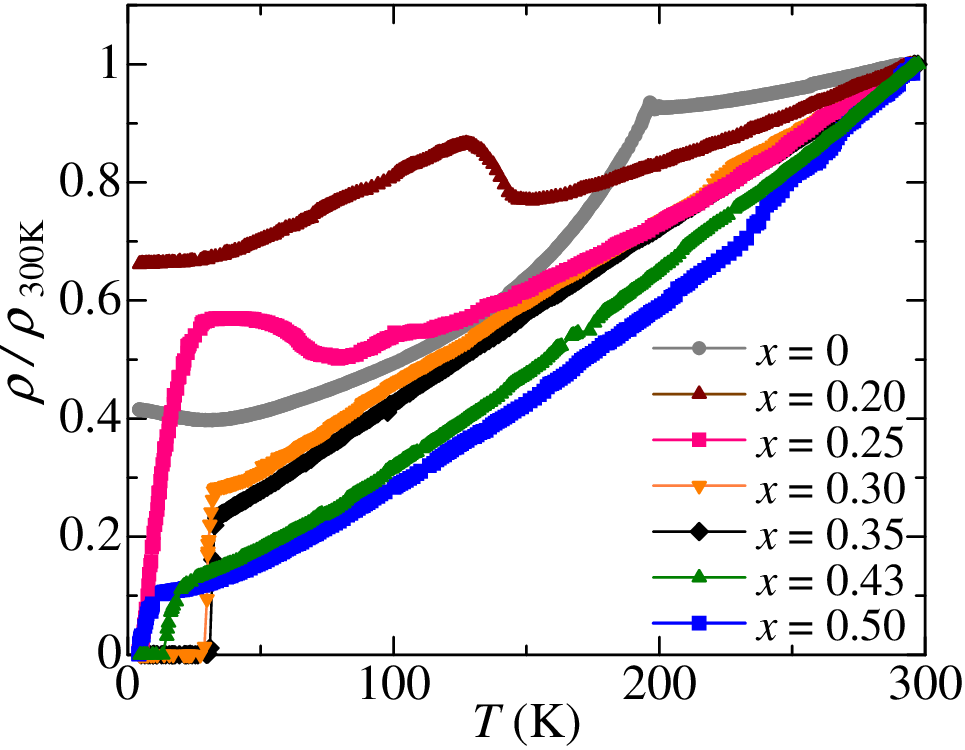}
\caption{(Color online)
Temperature dependence of the in-plane electrical resistivity of SrFe$_{2}$(As$_{1-x}$P$_{x}$)$_{2}$ for $x=0, \,0.20,\,0.25,\, 0.30,\,0.35,\,0.43,\,0.50$, normalized to the room temperature value.}
\label{resist}
\end{center}
\end{figure}

Figure \ref{resist} shows the temperature dependent in-plane electrical resistivity of SrFe$_{2}$(As$_{1-x}$P$_{x}$)$_{2}$ series, normalized to their room temperature value. For SrFe$_{2}$As$_{2}$, a sharp drop in resistivity at 197\,K is related to the structural and SDW transition. 
The upturn around 30\,K is due to the contamination of Sn. With P content increasing, the resistivity anomaly is suppressed and zero resistivity is attained at $x = 0.25$, indicating the coexistence of SDW and superconductivity. 
A superconducting temperature $T_c$ rises to 30\,K at $x = 0.35$. With more P substitution, $T_c$ is lowered to 20\,K at $x = 0.43$ and 10\,K at $x = 0.50$.
For $x = 0.35$, the resistivity exhibits $T$-linear dependence in a wide $T$ range which suggests that a non Fermi liquid like behavior governed by a magnetic quantum fluctuation. 
As $x$ further increases, the temperature dependence of resistivity changes towards $T^{2}$ which is consistent with a standard Fermi liquid behavior.  

\begin{figure}
\begin{center}
\includegraphics[width=8cm]{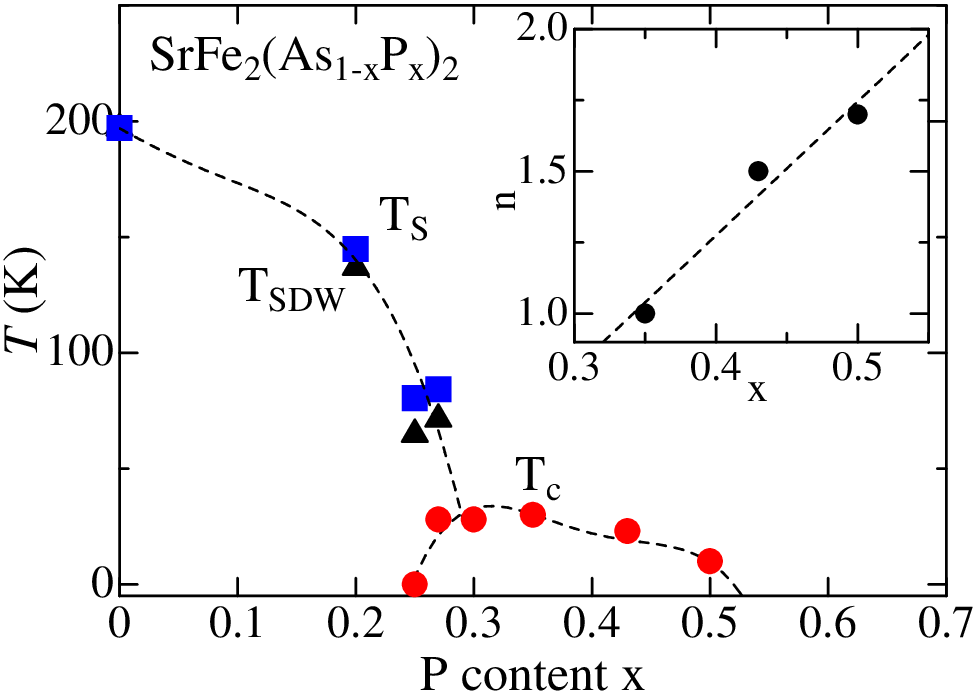}
\caption{(Color online)
$T$-$x$ phase diagram of SrFe$_{2}$(As$_{1-x}$P$_{x}$)$_{2}$ single crystals for $0\leq x \leq 0.50$.
$T_s$, $T_{\rm{SDW}}$ and $T_c$ are determined from the resistivity and susceptibility measurements.
The inset represents the power $n$ in resistivity fitted by $\rho(T)=\rho_{0}+AT^{n}$.}
\label{phase}
\end{center}
\end{figure}

Figure \ref{phase} displays the temperature-doping concentration ($T$-$x$) phase diagram obtained in this study. 
Structural, SDW and superconducting transition temperature were inferred from the resistivity and magnetic susceptibility measurements. 
Square symbols represent the structural transition temperature, $T_{s}$, while triangle symbols represent the magnetic transition temperature, $T_{\mathrm{SDW}}$. As it can be seen, the phase transition temperatures monotonically decrease as P content increases. 
For $x > 0.25$, a dome like superconducting phase appears, while the structural / magnetic transition disappears. The superconducting transition temperature, $T_c$, which is represented by circle symbols, reaches maximum value of 30\,K for $x = 0.30$ and $0.35$, then decreases to 20\,K at $x = 0.43$ and 10\,K at $x = 0.50$. 
The inset represents the exponent $n$ in resistivity fitted by $\rho(T)=\rho_{0}+AT^{n}$. The $n$ changes from 1 of the non Fermi liquid toward 2 of the Fermi liquid with increasing $x$. 
This change suggests that the antiferromagnetic quantum critical point exists around $x = 0.35$. 
In addition, the recent NMR measurement of our single crystal of SrFe$_{2}$(As$_{0.65}$P$_{0.35}$)$_{2}$ also found that it was close to the magnetic quantum critical point\cite{3}.

\section{SUMMARY}
We have synthesized the series of single crystals of SrFe$_{2}$(As$_{1-x}$P$_{x}$)$_{2}$ and studied the physical properties. 
The result of resistivity measurement revealed the pronounced non-Fermi-liquid like behavior at the SDW quantum critical point around $x=0.35$. 
This behavior is similar result of P substituted Ba122, suggesting that antiferromagnetic fluctuation plays an important role in the superconducting mechanism in this P substituted 122 system.

\section{ACKNOWLEDGEMENT}
This research was supported by Strategic International
Collaborative Research Program (SICORP), Japan Science and Technology
Agency.

\end{document}